\documentclass{aa} 
\usepackage{longtable}
\usepackage{txfonts} 
\usepackage{graphicx} 
\usepackage{natbib} 
\bibpunct{(}{)}{;}{a}{}{,} 

\def\tauCet{$\tau$~Cet}
\def\epsEri{$\epsilon$~Eri}
\def\delEri{$\delta$~Eri}
\def\alfHor{$\alpha$~Hor}
\def\alfHya{$\alpha$~Hya}
\def\alfMon{$\alpha$~Mon}
\def\zetPyx{$\zeta$~Pyx}
\def\lamPyx{$\lambda$~Pyx}
\def\delCrt{$\delta$~Crt}
\def\ksiHya{$\xi$~Hya}
\def\betVir{$\beta$~Vir}
\def\gamHya{$\gamma$~Hya}
\def\piHya{$\pi$~Hya}
\def\alfCenA{$\alpha$~Cen~A}
\def\alfCenB{$\alpha$~Cen~B}
\def\gamAps{$\gamma$~Aps}
\def\muAra{$\mu$~Ara}
\def\delPav{$\delta$~Pav}
\def\epsInd{$\epsilon$~Ind}

\begin{document} 

\title{Bisectors of the HARPS Cross-Correlation-Function\thanks{Based on observations collected at the La Silla Observatory, ESO (Chile), with the HARPS spectrograph at the ESO 3.6m telescope obtained from the ESO/ST-ECF Science Archive Facility.}} 
\subtitle{The dependence on stellar atmospheric parameters}

\author{
\"{O}. Ba\c{s}t\"{u}rk \inst{1,2} 
	\and 
T. H. Dall \inst{1} 
\and
R. Collet \inst{3}
\and
G. Lo Curto \inst{1}
\and
S. O. Selam\inst{4}}

\offprints{T. H. Dall, \email{tdall@eso.org}}

\institute{
European Southern Observatory, Karl-Schwarzschild-Str.~2, D-85748 Garching bei M{\"u}nchen, Germany
	\and 
Ankara University, Astronomy \& Space Sciences Research and Application Center, TR-06837 Ahlatlibel, Ankara, Turkey
\and
Max-Planck-Institut f{\"u}r Astrophysik, Karl-Schwarzschild-Str.~1, D-85741 Garching bei M{\"u}nchen, Germany
\and
Ankara University, Faculty of Science, Department of Astronomy and Space Sciences, TR-06100 Tandogan-Ankara, Turkey
}

\date{Received date/ Accepted date}

\abstract 
{Bisectors of the HARPS cross-correlation function (CCF) can discern between planetary radial-velocity (RV) signals and spurious RV signals from stellar magnetic activity variations. However, little is known about the effects of the stellar atmosphere on CCF bisectors or how these effects vary with spectral type and luminosity class.} 
{Here we investigate the variations in the shapes of HARPS CCF bisectors across the HR diagram in order to relate these to the basic stellar parameters, surface gravity and temperature.} 
{We use archive spectra of 67 well studied stars observed with HARPS and extract mean CCF bisectors. We derive previously defined bisector measures (BIS, $v_\mathrm{bot}$, $c_\mathrm{b}$) and we define and derive a new measure called the CCF Bisector Span (CBS) from the minimum radius of curvature on direct fits to the CCF bisector. } 
{We show that the bisector measures correlate differently, and non-linearly with $\log g$ and $T_\mathrm{eff}$. The resulting correlations allow for the estimation of $\log g$ and $T_\mathrm{eff}$ from the bisector measures. We compare our results with 3D stellar atmosphere models and show that we can reproduce the shape of the CCF bisector for the Sun.} 

\keywords{
Instrumentation: spectrographs --
Techniques: radial velocities --
Line: profiles --
Stars: atmospheres --
Stars: activity 
}
\maketitle

\section{Introduction}
Fundamental stellar parameters are best determined by studying high resolution, high signal-to-noise spectra. High resolution spectroscopy is also essential for studying velocity fields in stellar atmospheres. There are a number of astrophysical phenomena leading to the emergence of macroscopic velocity fields in stellar atmospheres, which can be observed as spectral line asymmetries and radial velocity variations. Line bisectors, defined as the the loci of the midpoints on the horizontal lines extending from one side to the other of spectral line profiles, are often employed to study the mechanisms that cause such asymmetries and  variations. \citet{gray2005} related line asymmetries to a fundamental stellar parameter by showing that the blue-most point of the FeI $\lambda$6252 line bisector is a very good indicator of luminosity class for the stars from late G to early K spectral types.

Although it is possible to obtain high-resolution spectra with the use of state of the art telescope-spectrograph configurations, in order to achieve sufficient signal-to-noise ratio (SNR hereafter), one typically has to either use long exposure times and/or combine many individual exposures.  With this approach, however,  short-term line profile variations cannot be measured. To overcome this limitation, one can either average bisectors of individual lines or employ cross-correlation functions (CCFs) instead. These functions are computed by cross-correlating observational stellar spectra with synthetic line masks with the objective of preserving line profile information; they can be thought of as near-optimal ``average" spectral lines because the cross-correlation is based on lines of many different elements \citep{dall+2006}. Because of this, CCFs have high SNR and represent one single instance of time. In addition, their computation is faster compared to averaging bisectors of many individual lines. On the other hand, averaging bisectors make it easy to select the individual lines according to line strength, excitation potential etc. However, one can also design cross-correlation masks by selecting lines with similar strength and/or excitation potential, or with a limited number of elements or ions. Designing masks other than the standard masks in the HARPS Data Reduction Software (DRS hereafter) is an ongoing study and will be the subject of another paper.

Line bisectors delineate line asymmetries. In order to quantify the shape of bisectors, measures are defined. Bisector measures frequently used in the literature are mainly based on the \emph{velocity span} defined by \citet{tonergray1988} as the velocity difference between two bisector points at certain flux levels of the continuum on a single line bisector, and the \emph{curvature} defined by \citet{grayhatzes1997} as the  difference of two velocity spans. Most of the extrasolar planet survey studies \citep{queloz+2001,martinezf+2005} take the definition of \citet{tonergray1988} as reference for the velocity span and adopt it to CCF bisectors using average velocities in two different regions on the bisector, one at the top and one close to the bottom instead of  two real bisector points as \citet{tonergray1988} did. They investigate if the variation of the bisector measure correlates well with that of the radial velocity. If such a correlation exists, then the radial velocity variation is attributed to stellar mechanisms instead of a planetary companion which would not cause variations in line asymmetry.

It is \citet{dall+2006} that first gave a quantitative relationship between the stellar parameters (surface gravity, and absolute magnitude) and a combination of two bisector measures, namely the \emph{bisector inverse slope} (BIS hereafter) and the curvature.  Although their relation is based on a few data points, they stress that it can be possible to obtain the luminosity of a star from the shape of its average CCF bisector. Hence, the CCF bisector measures can be used as indicators of luminosity similar to the blue-most point on classical single line bisectors as \citet{gray2005} has shown.

In this paper, we investigate the relationship between stellar parameters (specifically surface gravity and effective temperature), and mean CCF bisector shapes of a number of late type stars observed with HARPS. In order to do this, we make use of the bisector measures frequently used in the literature as well as a new measure introduced here to represent the relative position of the upper segments of CCF bisectors.

\section{Observations and data analysis}
The data presented here are obtained with the HARPS spectrograph \citep{harps2003} which is installed on the 3.6~m telescope at the La Silla Observatory. HARPS provides high spectral resolution (115000 on average) and it is very efficient in providing high radial velocity precision by keeping the instrumental profile (IP hereafter) stable thanks to 	its ultra-high internal stability.

We selected 67 late-type stars observed with HARPS (see Table~\ref{tab:obs}), for which stellar parameters have been published previously. In order to ensure a consistent set of stellar parameters across our data set, we preferably use references that determined these based on high-resolution spectra, and preferably obtained with HARPS.
We cover the entire region of the HR diagram from late F to late K. Unfortunately, there are not many M stars in the HARPS archive with good SNR, and which have previously published trustworthy stellar parameters obtained by high resolution spectroscopy. Therefore, our study is limited to the spectral range from late F to late K stars. Our data set covers a wide range of surface gravities. 

Activity indices (Mount Wilson S-index \citep{vaughan+1978} and $\log R'_{HK}$ \citep{noyes1984}) (see Table~\ref{tab:obs}) of all our program stars were determined with the DRS.

As a rule of thumb, we eliminated spectra with SNR lower than 150 in the $60^{th}$ ech{\'e}lle order ($\sim$ 6025-6085~\AA) as calculated by the DRS. We inspected CCFs visually to guard against any obvious problems. HARPS CCFs are calculated for each of the 72 echelle orders, which are then averaged and included by the DRS as a $73^{rd}$ order in the data products. We then extracted the bisectors of these averaged CCFs by taking the midpoints of horizontal lines connecting observation points on one wing to points on the other found by a cubic spline interpolation for each of the wings. We performed the procedure by using a new python code, developed by us. We obtained the radial velocities from the fits headers, that have been determined by the HARPS DRS from the centers of the Gaussian fits to the CCF profiles.

\begin{figure}
\resizebox{\hsize}{!}{
\includegraphics{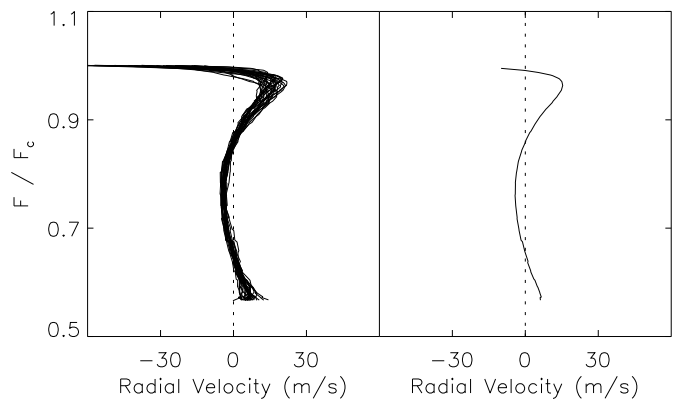}}
\caption{\label{fig:HR8658meanbis}\emph{Left:} The CCF bisectors of \object{HR 8658} (G0.5 V) computed using a G2 mask. \emph{Right:} The mean CCF bisector.}
\end{figure}

In order to maximize the SNR, the mean bisector for each star has been computed, an example of which is given in Fig.~\ref{fig:HR8658meanbis}. All the measurements presented in this paper have been performed on the mean bisector for each star. However, in order to show the potential of our method in determining stellar parameters from a single high-resolution spectrum, we point out that the minimum SNR in the observations of \object{HR 8658} ($V = 6^{m}.64$) is 160, which has been reached in 230 seconds of integration time.  For a $10^{m}.5$ star, it is possible to reach an SNR of 150 in a one hour exposure.

\begin{figure}
\resizebox{\hsize}{!}{
\includegraphics{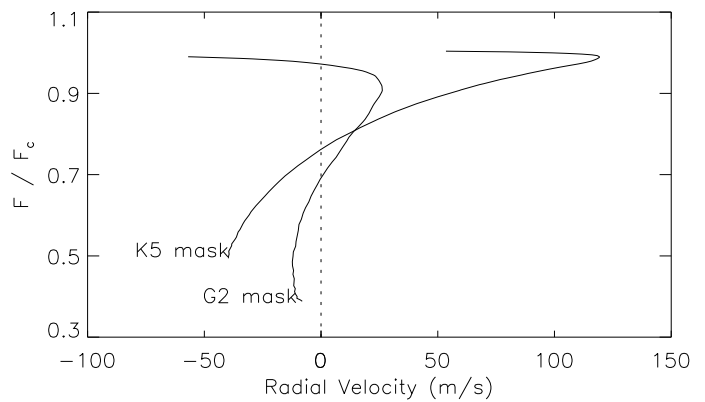}}
\caption{\label{fig:mask_alfHor}Mean bisectors of \object{\alfHor} (K2 III) obtained with G2 and K5 masks.}
\end{figure}

We used the CCFs computed by the HARPS pipeline (DRS). These functions are obtained by cross-correlating stellar spectra with a standard stellar mask matching as closely as possible the spectral type of the star. 
In Fig. \ref{fig:mask_alfHor}, mean CCF bisectors of \object{\alfHor} obtained by using the standard G2 and K5 masks of the HARPS DRS are given as an example.

\citet{dall+2006} demonstrated that the shape of the CCF depends strongly on the mask used in the cross-correlation procedure (cf.~their Fig.~3).  In effect, the resultant CCFs obtained with different masks will be ``averages" of different spectral lines. 
Thus, in order to take into account the effect of the mask when comparing CCF bisectors of different stars, we obtained mean CCF bisectors of all our program stars using both of the masks available in the HARPS DRS. This was accomplished during a data reduction mission to ESO Headquarters.

\subsection{Error Analysis}
The error on each bisector point has been computed by following the procedure given by \citet{gray1983}. The error in radial velocities of our bisector points coming from photometric error can be expressed as,

\begin{equation}\label{eq:error}
        \mathrm{\delta V_{i}} = \frac{1}{\sqrt{2}} \frac{1}{\mathrm{SNR \sqrt{1-\alpha_{i}}}} \frac{1}{\mathrm{(dF/dV)_{i}}}.
\end{equation}

where  $\mathrm{1/(SNR \sqrt{1-\alpha_{i}})}$ term represents the photometric error on flux for each bisector point i, where the line depth is indicated by $\alpha_{i}$, and $\mathrm{dF/dV}$ is the slope of the profile at each point from which a bisector point is computed. The slope is obtained for each point from a linear fit to an array of points on the CCF profile, including the point itself and two adjacent points to it from both sides. 

The error on each bisector point is propagated on the bisector measures in a standard fashion by adding quadratically the errors of the bisector points used in the computation of the measure.

\subsection{Consistency of Bisectors}
Here we will discuss spectrum-to-spectrum variations of CCF bisectors, and their effects on the computation of mean bisectors and bisector measures. The RV for each spectrum has been calculated by the HARPS DRS by a Gaussian fit to the CCF profile. This RV has been subtracted from the individual CCF bisector. As a result, spectrum-to-spectrum variations in the CCF bisector are mainly caused by
(1) SNR differences due to differences in exposure time and variations in the conditions of the Earth's atmosphere,
(2) instrumental effects, e.g. temperature drift, 
(3) errors coming from the reduction procedure,
(4) variations in the stellar atmosphere due to magnetic activity, stellar oscillations, granulation, and any other variable velocity fields. Note that orbiting planets do not leave a signature on the CCF bisectors.

\begin{figure}
\resizebox{\hsize}{!}{
\includegraphics{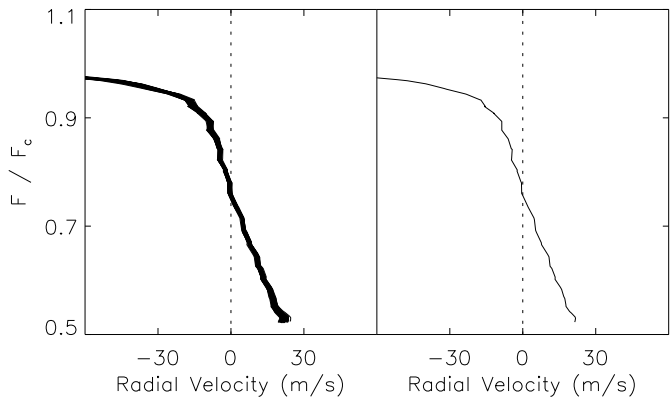}}
\caption{\label{fig:alfCenAmeanbis}\emph{Left:} The CCF bisectors of \object{\alfCenA} (G2 V) computed using a G2 mask. \emph{Right:} The mean CCF bisector. Note that the small scale structure of the bisector is constant and thus likely real.}
\end{figure}

In order to illustrate to what degree the bisectors are stable, we show the individual bisectors of \object{\alfCenA} obtained during more than 10 hours of asteroseismological observations in the night of 21 April 2005 as well as their mean (Fig.~\ref{fig:alfCenAmeanbis}). The SNR of the mean bisector is moderate ($\sim$875) compared to that for other stars in  Table~\ref{tab:obs}, all of which were observed for durations less than the total obervation time allocated for \object{\alfCenA} in one night.   

\begin{figure}
\resizebox{\hsize}{!}{
\includegraphics{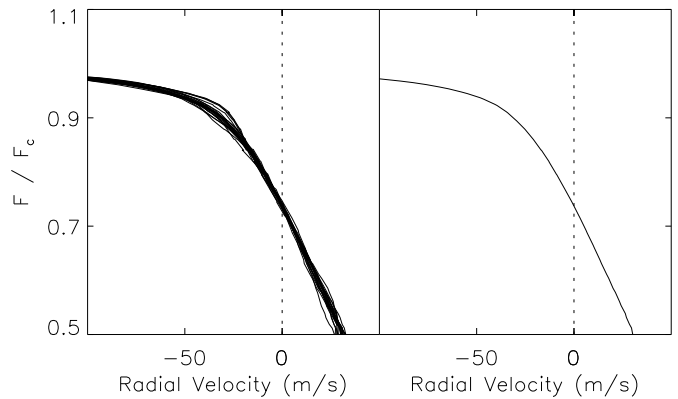}}
\caption{\label{fig:HD21693meanbis}\emph{Left:} The CCF bisectors of \object{HD 21693} (G9 IV-V) computed using a G2 mask. \emph{Right:} The mean CCF bisector.}
\end{figure}

To illustrate the long term consistency,  21 CCF bisectors of \object{HD 21693} obtained in 21 different nights in the period 26 December 2003 --- 1 September 2009 is shown in Fig.~\ref{fig:HD21693meanbis}. Fig.~\ref{fig:alfCenAmeanbis} and \ref{fig:HD21693meanbis} show that the IP is very stable, both short term and long term.

\begin{figure}
\resizebox{\hsize}{!}{
\includegraphics{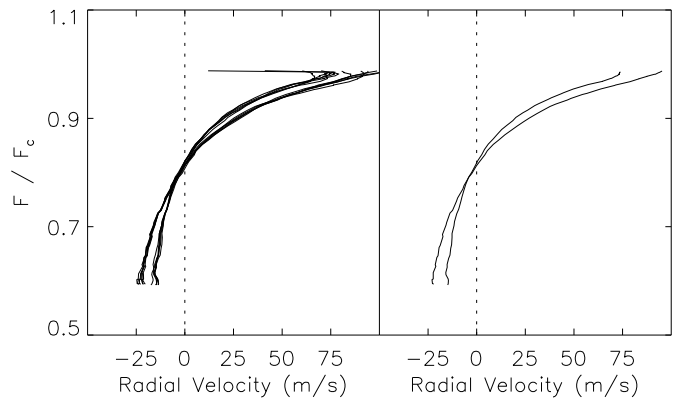}}
\caption{\label{fig:epsEri}\emph{Left:} Individual CCF bisectors of \object{\epsEri} (K2 V) computed using a K5 mask. \emph{Right:} The mean CCF bisectors for each night.}
\end{figure}

An interesting spectrum-to-spectrum variation was observed for \object{\epsEri}, which is a well-known active star \citep{croll+2006,graybaliunas1995}. The star was observed on two nights, on 25 January 2007 and 2 February 2007. Five spectra were obtained on each of the nights. The change in individual CCF bisectors from one night to another is remarkable (Fig.~\ref{fig:epsEri}), and illustrates both intra-night short term variation and longer term variations due to magnetic activity. In such cases, we used a mean CCF bisector obtained as an average of the bisectors observed in one night during which the shape is the most stable.   

\subsection{Quantitative Bisector Measures}
\label{measures}
A systematic change of CCF bisector shape with surface gravity and with effective temperature is expected. This was demonstrated by \citet{gray2005} for classical single line bisector shapes over the HR diagram. In Sec.~\ref{results} we present the equivalent for the HARPS CCF bisectors.
However, in order to quantify the variations, a measure that can numerically represent the bisector shape is required. 

\citet{gray2005}, made use of the height of the blue-most point of a single line  bisector as an empirical indicator of the luminosity of a star. However, CCF bisectors have shapes considerably different than those of single lines. The famous ``C" shape of the bisector of the neutral iron line at $\lambda$6253 in cool stars's spectra is physically attributed to granulation. Because granulation is depth dependent, the shape of the ``C", and hence its blue-most point, changes from one star to another depending on how transparent the atmosphere is to granulation. Therefore, such a measure referring to the height of that point would be related to the surface gravity of the stars in our sample. However, as it is clear from the bisectors shapes given in both HR diagrams, there is not always a blue-most point on CCF bisectors, due to their nature as ``average" spectral lines. In addition, if the HARPS IP is not symmetric, this may affect the shape of the bisector, and could cause a blue-most point to disappear. \citet{livingston+1999} defined \emph{bisector amplitude} as a proxy of the bisector shape to study line asymmetries observed in solar spectra. This measure also refers to an extreme blue point on a bisector. Furthermore, we will not be able to use the term \emph{flux deficit}, which was defined and proposed to be increasing with effective temperature and luminosity by \cite{gray2010} because the physical context on which the definition is based, depends very strongly on absolute positions of single spectral lines. 

\citet{dall+2006} made use of four bisector measures in their study on CCF bisectors. They adopted the definiton of velocity span as the difference of average velocities between 10-40\% and 55-90\% of the line depth from the top, which is the \emph{bisector inverse slope} \citep{queloz+2001}, leaving the term ``velocity span'' for classical single line bisectors. They also adopted a \emph{curvature} term, $c_b$, which they define in segments, 20-30\% (v$_1$), 40-55\% (v$_2$), and 75-100\% (v$_3$) of the line depth, as $c_b = $((v$_3$- v$_2$)-(v$_2$-v$_1$)), based on the definition of \citet{povich+2001}.  Additionally, they define the \emph{bisector bottom} as the average velocity of the bottom four points on the bisector, minus the RV of the star. In order to represent the straight segment on the central part of CCF bisectors, they define another bisector measure which they call \emph{bisector slope}. Because this measure is a variant of the BIS without any improvement, we will not use it.

\section{Results and Discussion}
\label{results}
\begin{figure}
\resizebox{\hsize}{!}{
\includegraphics{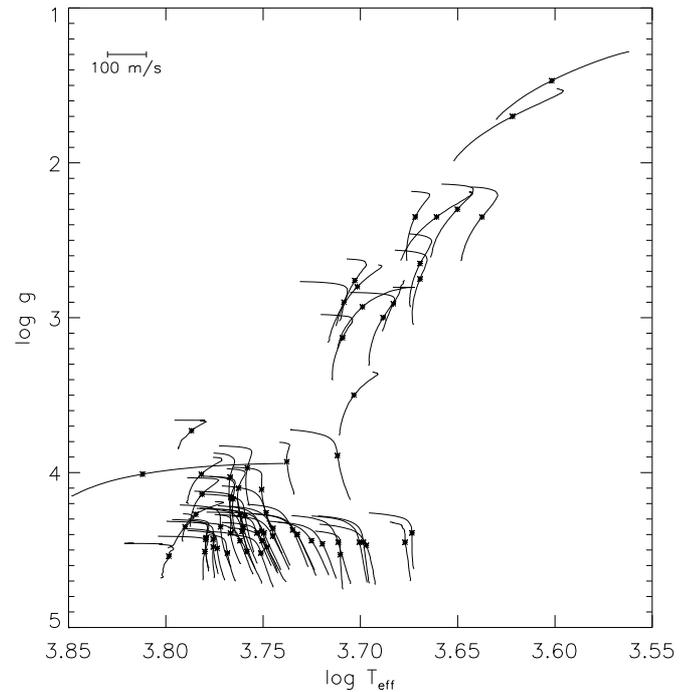}}
\caption{\label{fig:bisectorsHR_G2}Mean bisectors of program stars, computed by using a G2 mask, on an HR diagram. Asterisks on the half line depth show the position of the star on the HR diagram based on its parameters obtained from the literatue and given in Table~\ref{tab:obs}. The scale for radial velocities spanned by mean CCF bisectors is given on the upper left corner.}
\end{figure}

\begin{figure}
\resizebox{\hsize}{!}{
\includegraphics{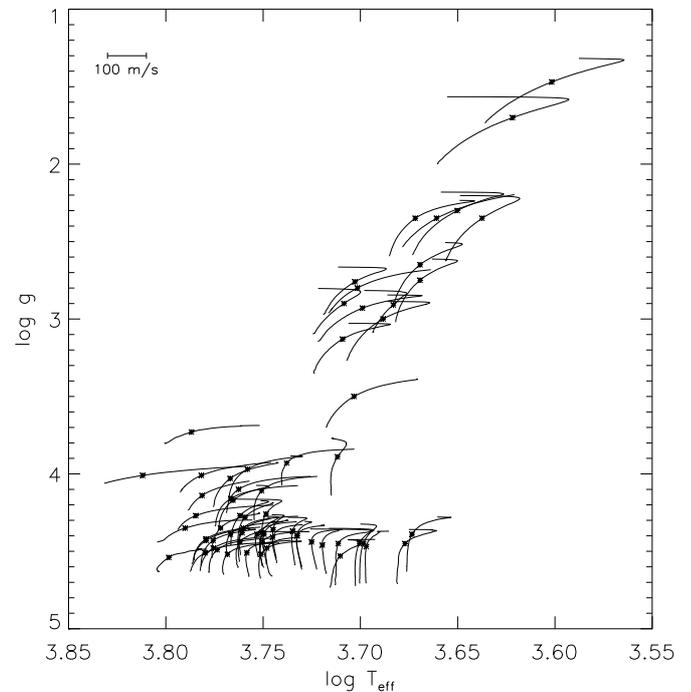}}
\caption{\label{fig:bisectorsHR_K5}Mean bisectors of program stars, computed by using a K5 mask, on an HR diagram. Same conventions apply as in Fig. \ref{fig:bisectorsHR_K5}.}
\end{figure}

In Fig.~\ref{fig:bisectorsHR_G2}, mean bisectors of our program stars obtained by using a G2 mask are illustrated on an HR diagram, based on their stellar parameters (Table~\ref{tab:obs}) depicted by asterisks on the half of the line depth. Mean bisectors on the HR diagram obtained by using a K5 mask are shown in Fig.~\ref{fig:bisectorsHR_K5} on the HR diagram with the same conventions as in Fig.~\ref{fig:bisectorsHR_G2}. All measurements were obtained by using the standard HARPS G2 and K5 masks, respectively

\subsection{Correlations Between Mean Bisector Measures and Stellar Parameters}  
\begin{figure}
\includegraphics{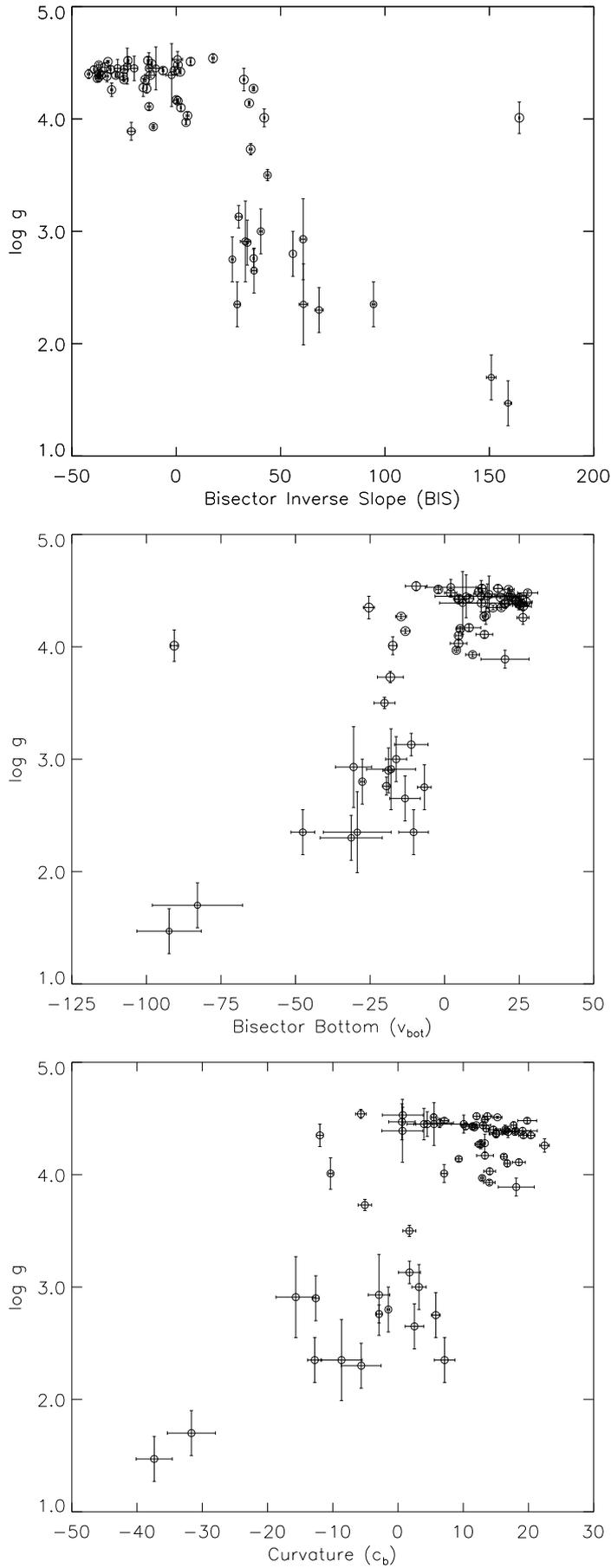}
\caption{\label{fig:corlogg_G2}Surface gravity vs. bisector measures based on mean bisectors obtained by using a G2 mask. Larger circle symbols indicate larger effective temperatures.}
\end{figure}

\begin{figure}
\includegraphics{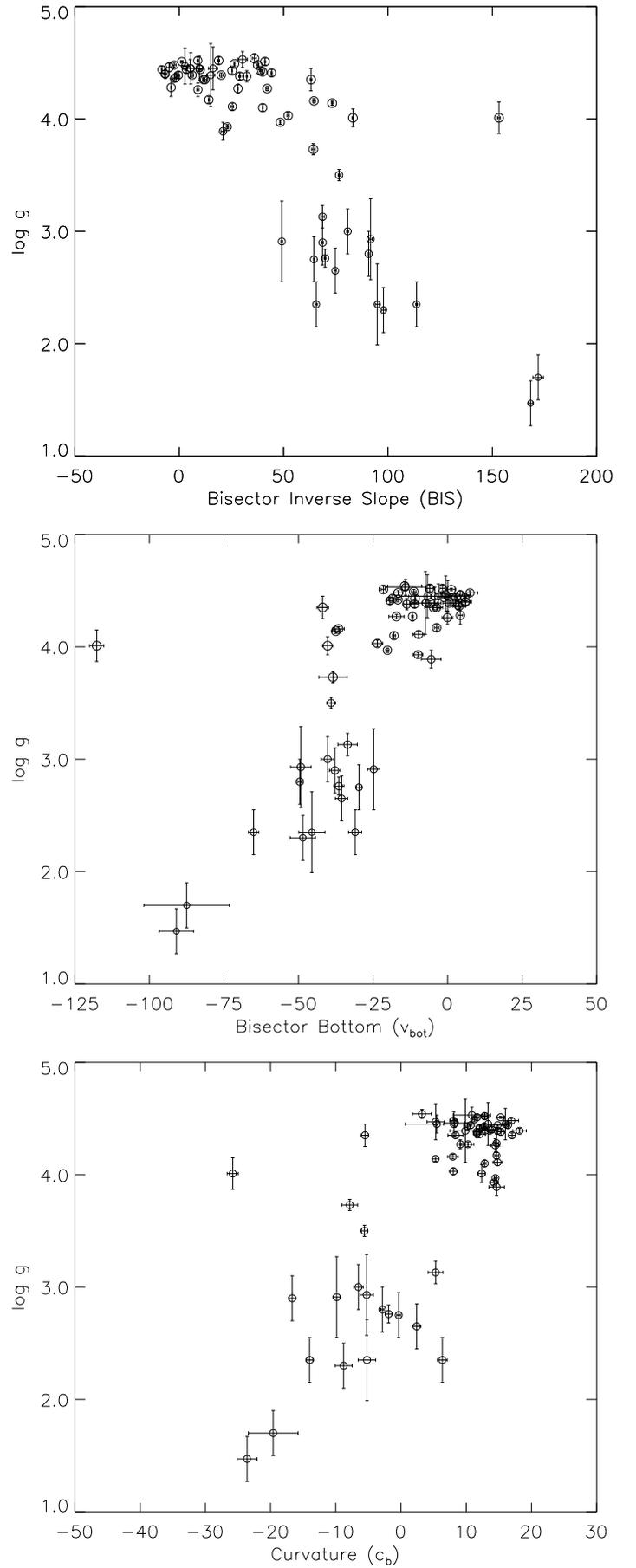}
\caption{\label{fig:corlogg_K5}Surface gravity vs. bisector measures based on bisectors obtained by using a K5 mask. Larger circle symbols indicate larger effective temperatures.} 
\end{figure}

A correlation between the bisector measures defined in Sec.~\ref{measures} and surface gravity is obvious in Figs.~\ref{fig:corlogg_G2} and~\ref{fig:corlogg_K5}. While BIS and v${_{bot}}$ both correlate well with surface gravity, c${_{b}}$ is less convincing.

\citet{dall+2006} gave a preliminary relationship between BIS+c$_{b}$ and the surface gravity.  Their relation is however based on only 9 solar type stars and biased by the effect of using both the G2 and the K5 masks. 

\begin{figure}
\includegraphics{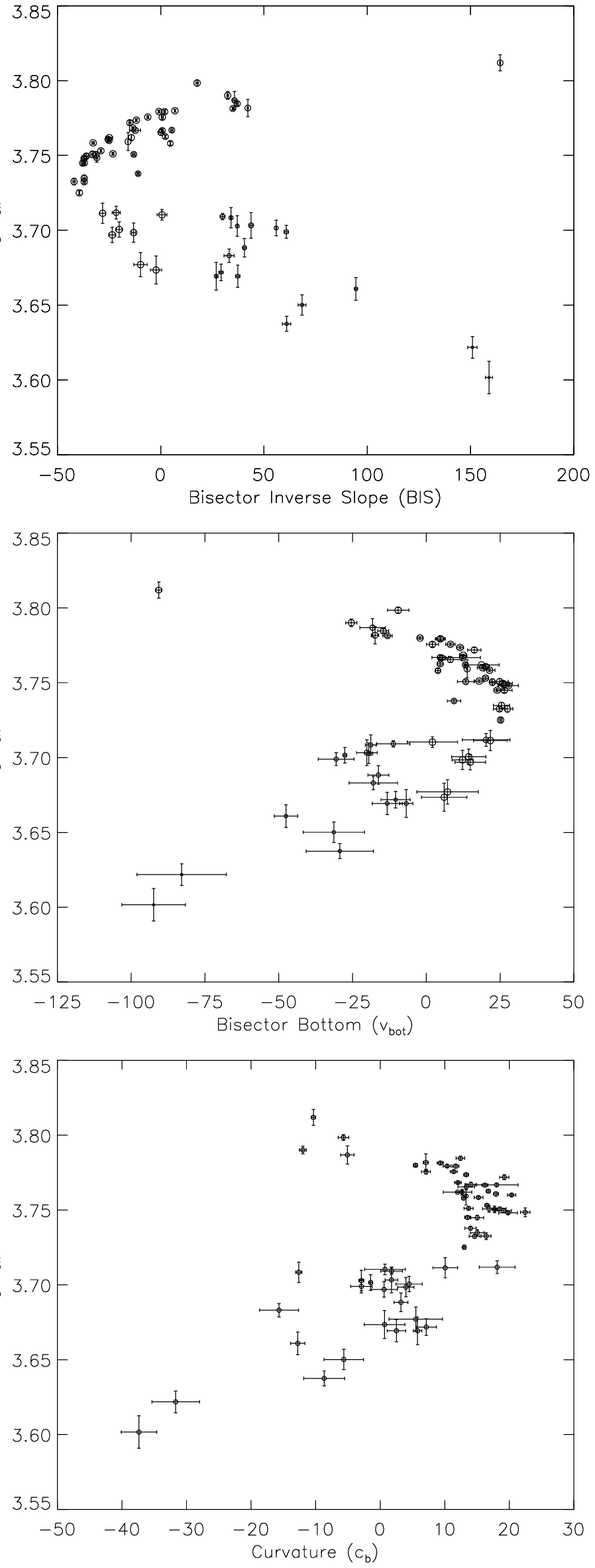}
\caption{\label{fig:corTeff_G2} 
Effective temperature vs. bisector measures obtained by using a G2 mask. Larger circle symbols indicate larger surface gravities.}
\end{figure}

\begin{figure}
\includegraphics{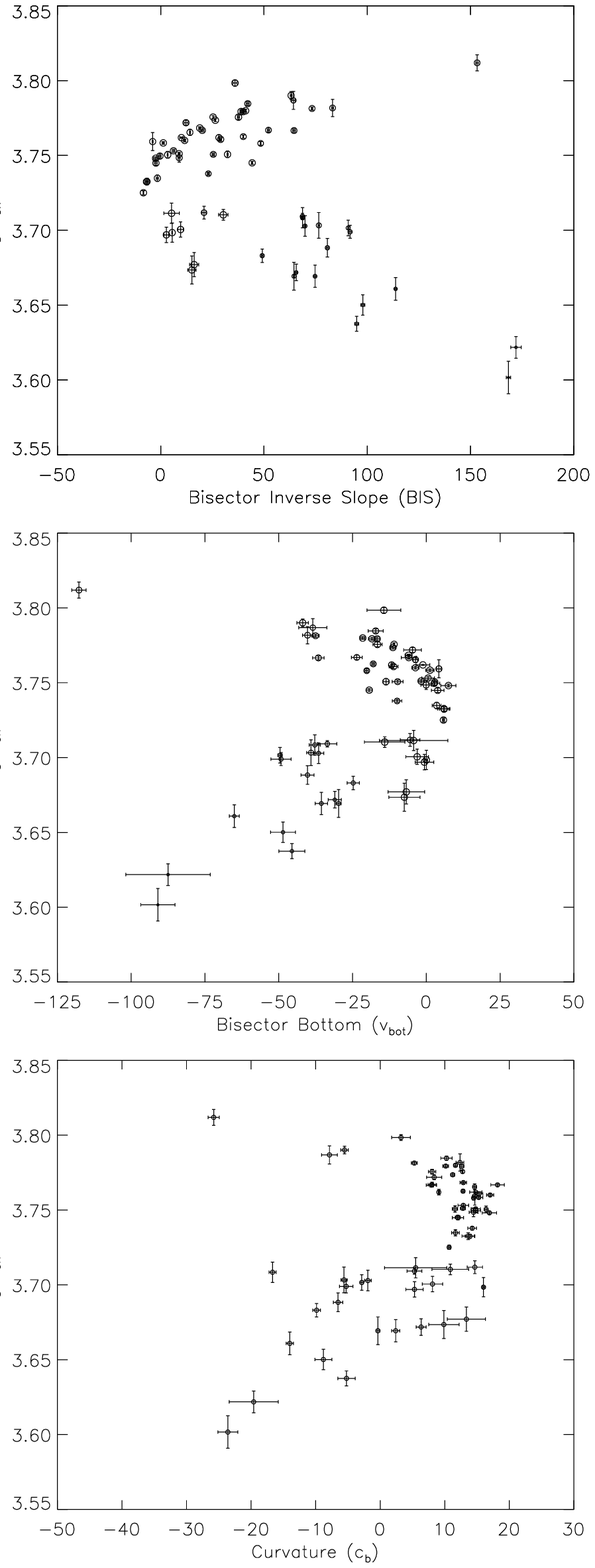}
\caption{\label{fig:corTeff_K5}Effective temperature vs. bisector measures obtained by using a K5 mask. Larger circle symbols indicate larger surface gravities.}
\end{figure}

The bisector measures do not correlate with effective temperature as linearly as they do with surface gravity (Fig. \ref{fig:corTeff_G2} for bisectors obtained by using a G2 mask, and Fig. \ref{fig:corTeff_K5} for bisectors obtained by using a K5 mask), especially for the dwarf stars in the sample. The star, for which the bisector measures deviate most in the correlations, is \object{Procyon}. It is by far the hottest star in our sample. Unfortunately, we do not have intermediate temperature stars between Procyon and the next hottest star in the sample to trace the  ``non-linear" correlation hinted at in both Fig. \ref{fig:corTeff_G2} and in Fig. \ref{fig:corTeff_K5}.

It is important to notice that the shape of the bisector also depends on the IP. Therefore, the correlations given here are valid only for the HARPS instrument and cannot be used to estimate stellar parameters for other instruments.

\subsection{Direct fitting of the bisector and new quantitative measures}
\begin{figure}
\includegraphics{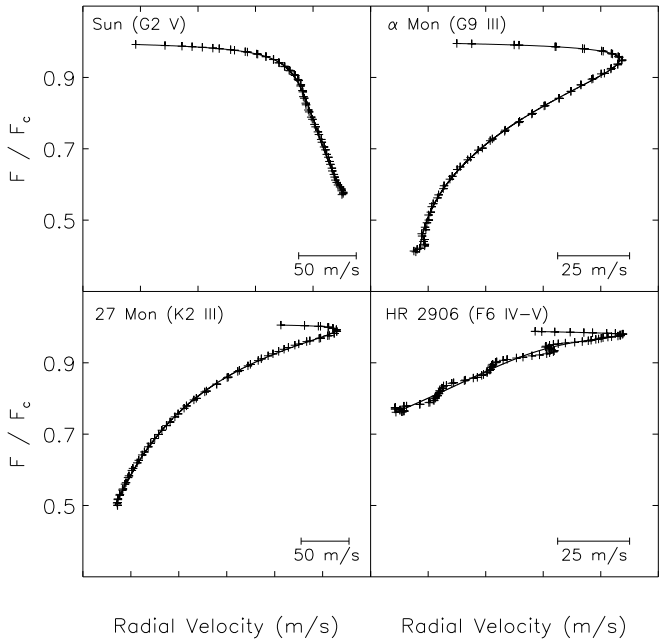}
\caption{\label{fig:nistfit} Mean CCF bisectors (plus signs) and appropriate NIST-Hahn fit functions (continuous curves) for the Sun, \object{27 Mon}, \object{HR 2906}, and \object{\alfMon} from topleft in counter-clockwise direction.} 
\end{figure}

In order to derive a bisector measure that is as objective as possible, we have attempted to fit the bisector with a general analytical function. We have found that it is possible to make use of NIST-Hahn functions \citep{hahn1979} for fitting the entire bisector for a wide variety of shapes (Fig.~\ref{fig:nistfit}). NIST-Hahn functions have the form

\begin{equation}\label{eq:nist}
        \mathrm{f(x)} = \frac{a+b\mathrm{x}+c\mathrm{x^{2}}+d\mathrm{x^{3}}}{1+f\mathrm{x}+g\mathrm{x^{2}}+h\mathrm{x^{3}}}
\end{equation}

where $x$ is the flux, and $f(x)$ is the velocity on the fit for a CCF bisector. Coefficients for the best fit for each mean CCF bisector has been obtained by Levenberg-Marquardt least-squares minimization by using the python module MPFIT \citep{markwardt2009}.

From the bisector shapes on the HR diagrams given in Figs. \ref{fig:bisectorsHR_G2} and \ref{fig:bisectorsHR_K5}, it is expected that the relative position of the upper segments of CCF mean bisectors changes systematically over the HR diagram. However, it is not possible to represent upper segments of CCFs with a single velocity value as the bottom velocity represents the lower segment. Because the entire bisector can be fit by a NIST-Hahn function, the radial velocity value at a point on that function where the radius of curvature has a minimum can be used as a measure of the relative position of the top of the bisectors. NIST-Hahn functions fit the overall bisector shape but not irregularities on bisectors. This is advantageous especially for ``wiggly" bisectors of F-type stars for which there is usually more than one point where the radius of curvature is low. Using NIST-Hahn functions; therefore, makes it possible to find a single point on the top of the bisectors where the radial velocity has a marginal value representing the top segment. 

Radius of curvature on a curve y, with one parameter x is given as

\begin{equation}\label{eq:radcurv}
        \mathrm{R} = \frac{|1+(\frac{dy}{dx})^2|^{3/2}}{|\frac{d^2y}{dx^2}|}.
\end{equation}

\begin{figure}
\resizebox{\hsize}{!}{
\includegraphics{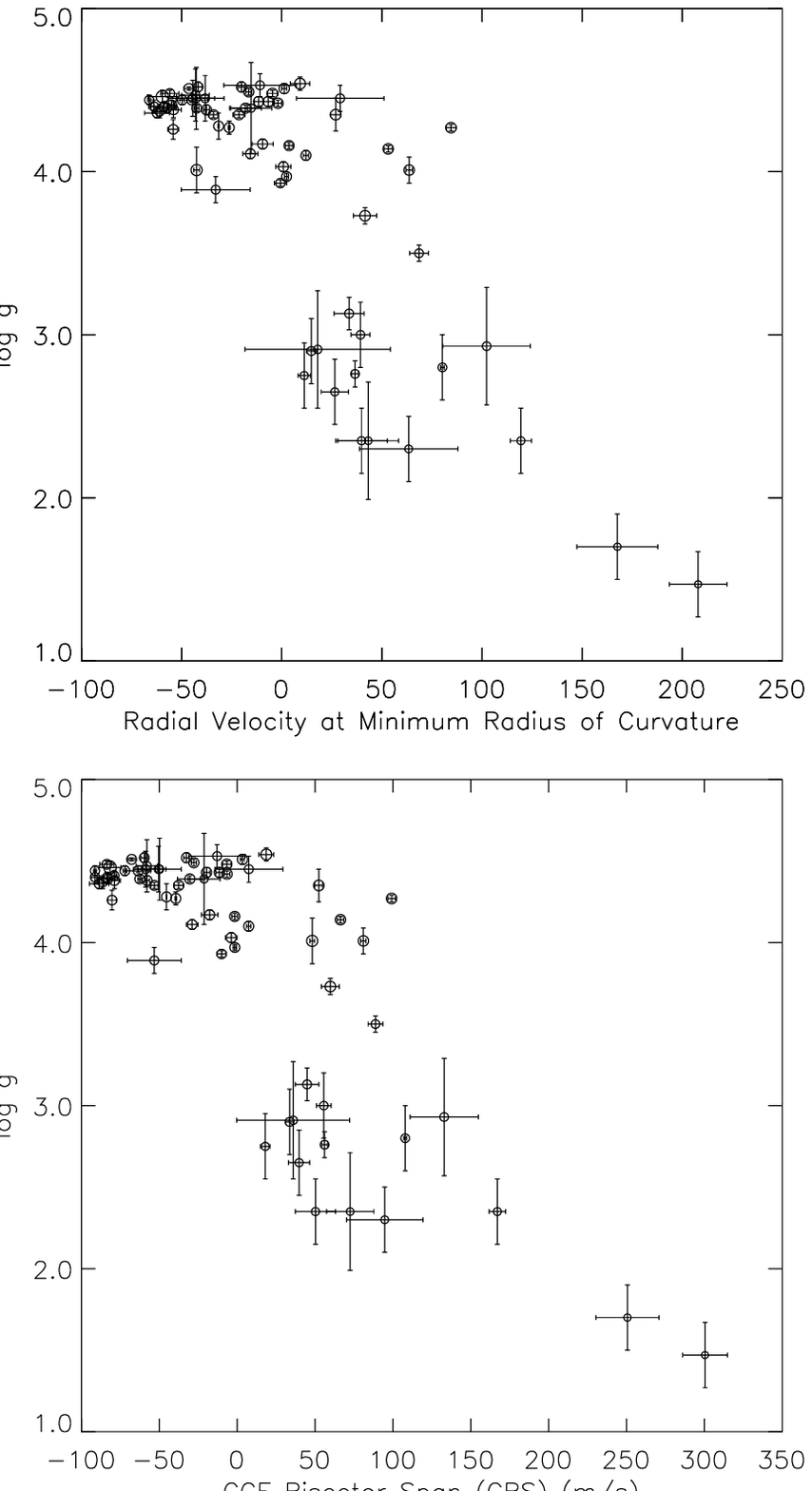}}
\caption{\label{fig:corlogg_radcurv_G2}\emph{Top:} Surface gravity vs. radial velocity at the minimum radius of curvature. \emph{Bottom:} Surface gravity vs. velocity span. Both measures are obtained on mean CCF bisectors which are computed by using a G2 mask. Larger circle symbols indicate larger effective temperatures.}
\end{figure}

\begin{figure}
\resizebox{\hsize}{!}{
\includegraphics{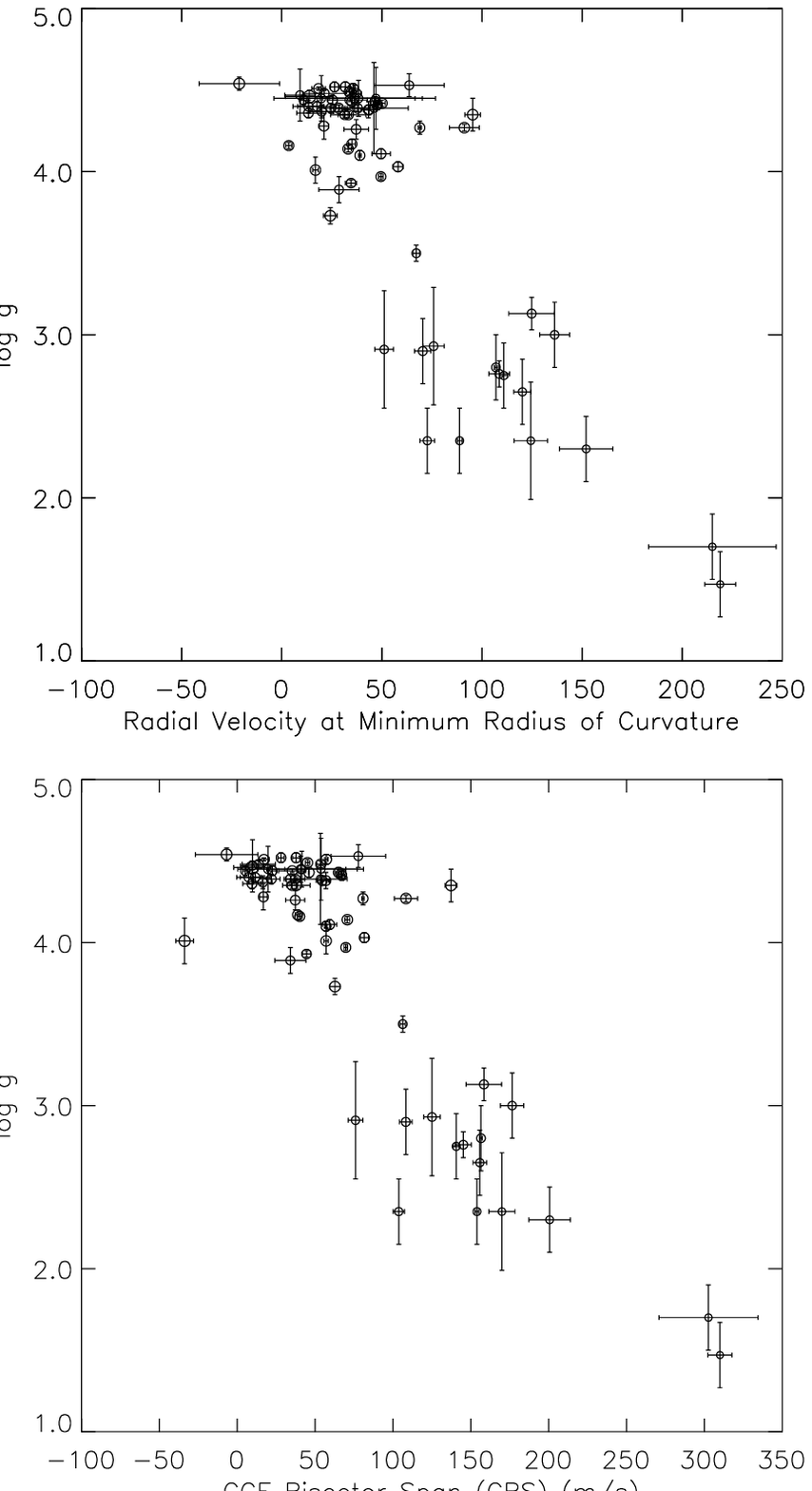}}
\caption{\label{fig:corlogg_radcurv_K5}\emph{Top:} Surface gravity vs. radial velocity at the minimum radius of curvature. \emph{Bottom:} Surface gravity vs. velocity span. Both measures are obtained on mean CCF bisectors which are computed by using a K5 mask. Larger circle symbols indicate larger effective temperatures.}
\end{figure}

\begin{figure}
\resizebox{\hsize}{!}{
\includegraphics{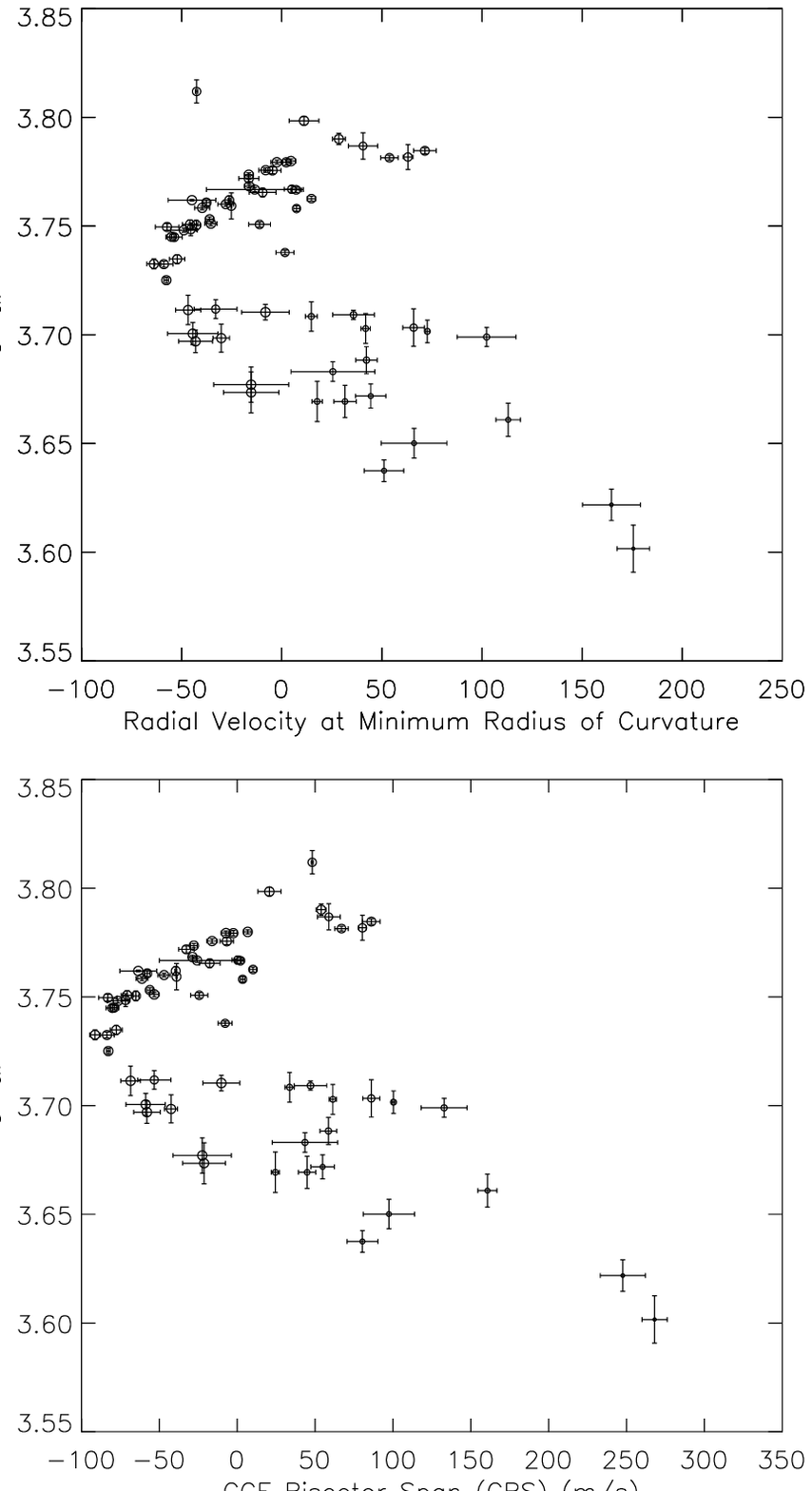}}
\caption{\label{fig:corTeff_radcurv_G2}\emph{Top:} Effective temperature vs. radial velocity at the minimum radius of curvature. \emph{Bottom:} Effective temperature vs. velocity span. Both measures are obtained on mean CCF bisectors which are computed by using a G2 mask. Larger circle symbols indicate larger surface gravities.}
\end{figure}

\begin{figure}
\resizebox{\hsize}{!}{
\includegraphics{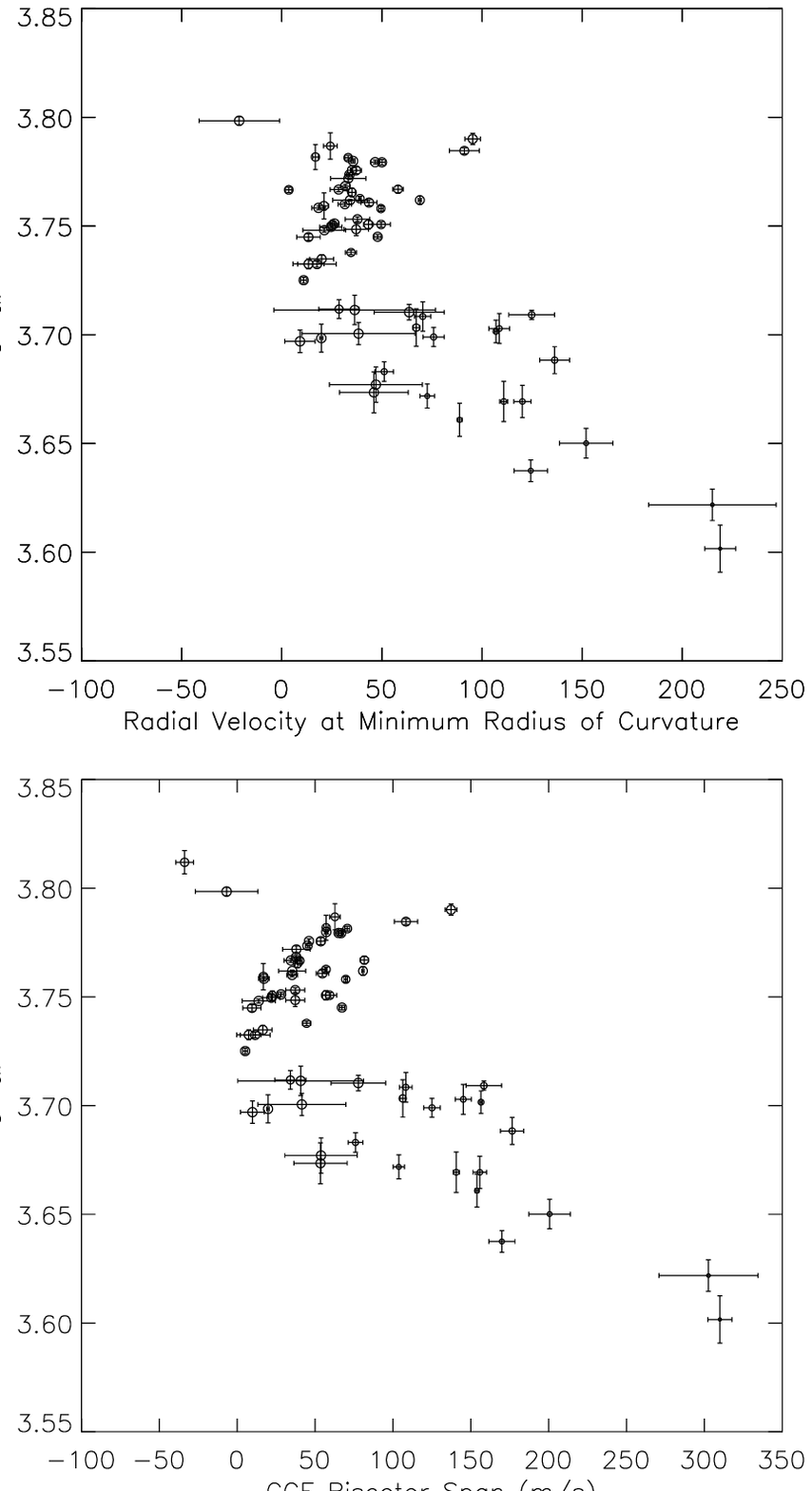}}
\caption{\label{fig:corTeff_radcurv_K5} \emph{Top:} Effective temperature vs. radial velocity at the minimum radius of curvature. \emph{Bottom:} Effective temperature vs. velocity span. Both measures are obtained on mean CCF bisectors which are computed by using a K5 mask. Larger circle symbols indicate larger surface gravities.}
\end{figure}

 We used Eq.~\ref{eq:radcurv} to analytically derive the radius of curvature at each point of the fit function (Eq.~\ref{eq:nist}) and find the velocity where the radius of curvature is minimum.  We used this velocity value as a bisector measure and invesigated if this measure is correlated with stellar parameters. Fig. \ref{fig:corlogg_radcurv_G2} and Fig. \ref{fig:corlogg_radcurv_K5} show that this new bisector measure is a good indicator of the surface gravity. Similarly, the correlation between this measure and effective temperature is obvious in Fig. \ref{fig:corTeff_radcurv_G2} and in Fig. \ref{fig:corTeff_radcurv_K5} for both masks used in the cross-correlation procedure. Hence a positon of a star on the HR diagram is related to the radial velocity corresponding to the point where the radius of curvature of its bisector has a minimum. 

It is hinted in the bisector shapes in Fig. \ref{fig:bisectorsHR_G2} and in Fig. \ref{fig:bisectorsHR_K5} that, not only the positions of the upper and lower segments of mean CCF bisectors but also the radial velocity range spanned by them  is related with the surface gravity and effective temperature. Bisector Inverse Slope (BIS) is a variant of the velocity span term defined by \citet{tonergray1988}. Hence it is a measure of the velocity spanned by CCF bisectors in addition to its characteristic as a measure of the slope. However, BIS is based on the difference between average velocities between certain flux levels, which do not always represent the top and bottom segments. Now that we have a good representation of the marginal velocities at the bottom with the \emph{bisector bottom} and at the top with the \emph{RV at the minimum radius of curvature}, we can define a good measure of velocity, that is spanned by mean CCF bisectors as the difference of these two measures. Since the term \emph{velocity span} is reserved for classical line bisectors mostly \citep{dall+2006}, we name this measure as \emph{CCF Bisector Span (CBS)}. We do not make use of the radial velocities at the bottom segment of the NIST-Hahn function because the points at the bottom of bisectors are not well constrained due to low SNR and incompetency of the interpolation function. Instead, we use the average velocity of four of them (bisector bottom). Fig. \ref{fig:corlogg_radcurv_G2}, \ref{fig:corlogg_radcurv_K5}, \ref{fig:corTeff_radcurv_G2}, and \ref{fig:corTeff_radcurv_K5} show that the CCF Bisector Span (CBS) is correlated very well with both surface gravity and effective temperature. 

\subsection{Comparisons with Synthetic Spectra}

In order to better understand the properties and asymmetries of CCFs, we carried out a comparison of the observed CCF bisectors with theoretical ones which we extracted from synthetic spectra. 
More specifically, we computed synthetic solar spectra for the regions $5150-5200$~{\AA} and $6215-6275$~{\AA} using a realistic, three-dimensional (3D), time-dependent, radiative-hydrodynamic simulation of solar surface convection.
The simulation domain is a rectangular domain covering an area of about $6{\times}6$~Mm horizontally and extending for about $7$~and $6$~pressure scale heights above and below the Sun's optical surface, respectively. 
The simulation was generated with the {\sc Stagger-Code} \citep{nordlund95}, at a numerical resolution of $240^3$. 
For the simulation, we employed a realistic equation-of-state  \citep{mihalas88}, continuous opacity data from \citet{gustafsson75} and Trampedach (2011, in prep.), and line opacity data from B. Plez \citep[priv. comm.; see also][]{gustafsson08}. For the chemical composition, we assumed the solar element mixture from \citet{agss09}.

We used the code {\sc SCATE} \citep{Hayek:2011} to carry out a full spectral synthesis of the two spectral regions indicated above. 
For the calculations, we adopted a list of spectral lines with parameters extracted from {\sc VALD} \citep{Piskunov:1995,Ryabchikova:1997,{Kupka:1999}}. 
We selected a sample of $20$~solar simulation snapshots taken at regular intervals and covering in total one hour of simulated solar time. 
For each snapshot, we solved the monochromatic radiative transfer equation under the approximation of local thermodynamic equilibrium (LTE) for ${\ga}5\,000$ wavelength points in each spectral region, for all grid points at the surface of the simulation, and along $16$ inclined directions.
Doppler shifts and broadening due to the presence of atmospheric velocity fields in the simulation are properly accounted for by the code, without the need to introduce additional micro- or macro-turbulence parameters. 
To model rotational broadening, however, we assumed a value of $v\,\sin{i}=2$~km/s for the Sun's rotational velocity.
The final flux profiles were computed by averaging the spectra over time and over all grid-points at the surface of the simulation and by performing a disc-integration over the solid angle. 
Prior to the analysis, the spectra were also rebinned and brought to the same wavelength resolution as the HARPS data.

Our observed spectra of the Sun have been obtained from the daytime sky. We extracted the above mentioned wavelength regions from one of our high SNR spectra and normalized the fluxes in these wavelength intervals to the continuum level by using {\sc IRAF}'s standard packages. We then cross-correlated both the synthetic and observed solar spectra with two different masks (``G'' and ``K''), specially designed for the same wavelength intervals. The masks available in the HARPS DRS pipeline are proprietary and not available outside the DRS, so they can only be used for the reduction of HARPS spectra. Hence, for these calculations, we decided to create new masks using {\sc VALD} line lists for the spectral regions under study. For the sake of simplicity, we used the ``extract stellar'' function from {\sc VALD} to retrieve a list of lines with estimated central depths expressed as a fraction of continuum flux for two generic solar-metallicity stars belonging to the two spectral types. We constructed the masks by simply overlapping a series of impulse functions centered at the wavelength of each spectral line and with amplitude proportional to the spectral line's depth. We were then able to cross-correlate both our synthetic and our observed spectra with the same masks. The cross-correlation was done using {\sc IDL}'s  {\sc c$\_$correlate} function. Finally, we extracted the bisectors from the resultant CCFs of synthetic and observed spectra and compared them with one another. 

\begin{figure}
\resizebox{\hsize}{!}{
\includegraphics{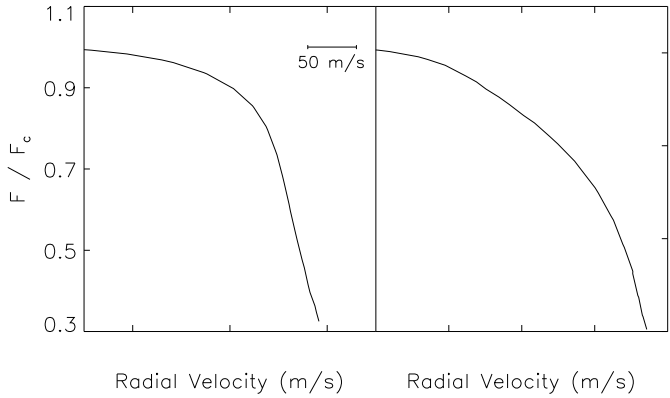}}
\caption{\label{fig:compbis_6215_6275_G8}\emph{Left:} The bisector of the CCF obtained by the cross-correlation of the solar synthetic spectrum and ``G'' mask for the $6215-6275$~{\AA} region. \emph{Right:} The bisector of the CCF obtained by the cross-correlation of a solar observed spectrum and ``G'' mask for the same region. Scale for x-axis is given on the figures and valid for both panels.}
\end{figure}

\begin{figure}
\resizebox{\hsize}{!}{
\includegraphics{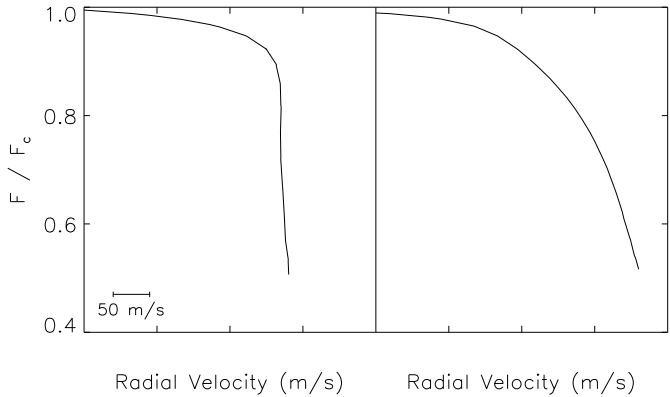}}
\caption{\label{fig:compbis_6215_6275_K9}\emph{Left:} The bisector of the CCF obtained by the cross-correlation of the solar synthetic spectrum and ``K'' mask for the $6215-6275$~{\AA} region. \emph{Right:} The bisector of the CCF obtained by the cross-correlation of a solar observed spectrum and ``K'' mask for the same region. Scale for x-axis is given on the figures and valid for both panels.}
\end{figure}

Fig.~\ref{fig:compbis_6215_6275_G8} shows the CCF bisectors extracted from the synthetic (left panel) and observed (right panel) spectra in the $6215$--$6275$~{\AA} range. The CCFs were computed using the ``G'' mask, that is the more closely corresponding to the Sun's spectral type. 
The agreement between the CCF bisectors from synthetic and observed spectra is very good: the curvature and asymmetry of the observed bisector are well reproduced by the theoretical calculations. 
Fig.~\ref{fig:compbis_6215_6275_K9} shows the CCF bisectors from the synthetic and observed spectra for the same wavelength interval computed using the ``K'' mask instead. While the curvature of the CCF bisector is more pronounced in the case of the synthetic spectrum,  the overall agreement with the observed case is still good. Small differences of these kind are acceptable considering, e.g., that lines may be missing from the list used for the synthetic spectrum calculations and that elemental abundances were not adjusted to reproduce all line strengths.

\begin{figure}
\resizebox{\hsize}{!}{
\includegraphics{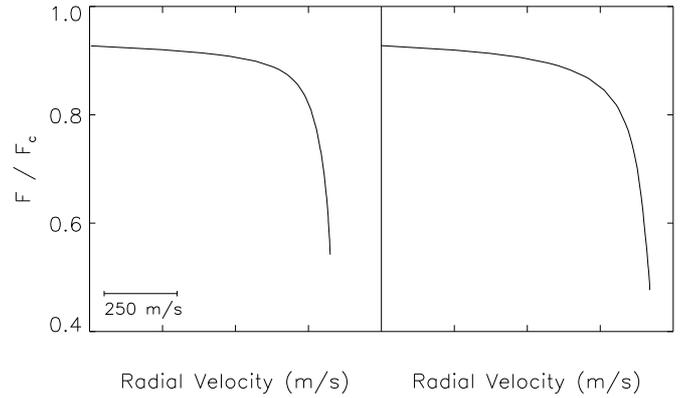}}
\caption{\label{fig:compbis_5150_5200_G8_mgincluded}\emph{Left:} The bisector of the CCF obtained by the cross-correlation of the solar synthetic spectrum and ``G'' mask for $5150-5200$~{\AA} region. \emph{Right:} The bisector of the CCF obtained by the cross-correlation of a solar observed spectrum and the same mask for the same region. The scale for x-axis is given on the figures and valid for both panels.}
\end{figure}

\begin{figure}
\resizebox{\hsize}{!}{
\includegraphics{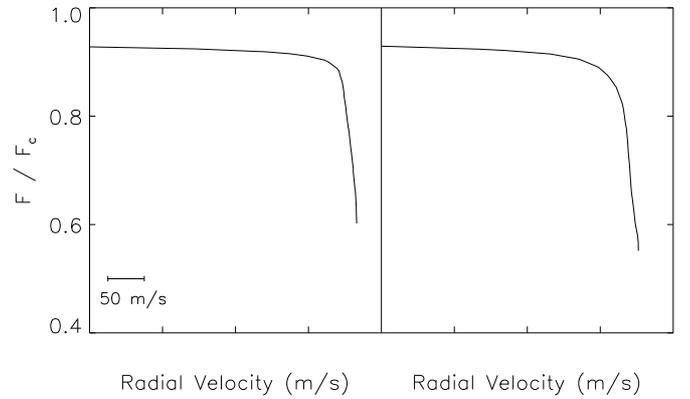}}
\caption{\label{fig:compbis_5150_5200_K9_mgincluded}\emph{Left:} The bisector of the CCF obtained by the cross-correlation of the solar synthetic spectrum and ``K'' mask for the $5150-5200$~{\AA} region. \emph{Right:} The bisector of the CCF obtained by the cross-correlation of a solar observed spectrum and the same mask for the same region. The scale for x-axis is given on the figures and valid for both panels.}
\end{figure}

The corresponding CCF bisectors for the $5150-5200$~{\AA} interval computed for the synthetic and observed spectra with the ``G'' and ``K'' masks are shown in Fig.~\ref{fig:compbis_5150_5200_G8_mgincluded} and~\ref{fig:compbis_5150_5200_K9_mgincluded}, respectively. 

This region is characterized by the presence of one moderately strong and two strong \ion{Mg}{i} lines at $5183$, $5172$,  and $5167$~{\AA} which carry a significant weight in determining the overall shape of CCF bisector in this region.  The agreement between CCF bisectors derived from synthetic and observed spectra is actually excellent for this region, implying that the modelling of these strong lines in our theoretical calculations is robust and satisfactory.

\section{Conclusions}
We have shown that the well known CCF bisector parameters BIS, $v_\mathrm{bot}$, and $c_b$ correlate well with $\log g$ and $T_\mathrm{eff}$.
We have constructed a new CCF bisector measure, the CBS, based on the point of least radius of curvature on the bisector, or alternatively, the point where  the rate of change of the velocity fields is largest. The CBS measure too correlates very well with $\log g$ and $T_\mathrm{eff}$. Although these correlations are not linear across the full range, we have shown that the shape of the CCF bisector changes in an understandable and predictable way across the HR diagram.

We have shown that current state-of-the-art 3D stellar atmosphere models are able to reproduce the behavior of the CCF quite well for the Sun, which is very encouraging. Further modeling will be needed in order to extend this understanding to other late type stars across the HR diagram.

The measures presented here have obviously predictive power; based on the CCF bisector, the fundamental atmospheric parameters of the star can be estimated. This may be relevant for studies of fainter stars where the SNR of individual spectral lines is not high enough for analysis. Such cases are still quite rare, partly because of the bias towards the study of brighter stars with HARPS.  Nevertheless, the results of this study can be useful in follow-up observations of exoplanet candidates discovered in transit surveys, by determining the parameters of especially the faint targets in a single observation based on the shape of its CCF bisector.

So far, most planet searches have been trying to smear or average out the effects of the stellar atmosphere \citep[e.g.,][]{dumusque+2011a,dumusque+2011b}. Only recently \citep{boisse+2011} has attempts been made to try to understand fully the effects of the stellar atmosphere with the aim of disentangling the effects. This is of particular importance in the context of the coming planet finding instruments, where sub-m/s precision will call for ways to discern and ultimately disentangle the effects of the stellar atmosphere from the planetary signal \citep[e.g., ESPRESSO for the VLT and later Codex for the E-ELT;][]{espresso,codex}. 

\begin{acknowledgements}
OB would like to thank The Scientific and Technological Research Council of Turkey (T\"{U}B\.{I}TAK) for the support by B\.{I}DEB-2211 scholarship and the ESO Director General for DGDF funding. 
TD and OB thanks ESO for kind support during a Data Reduction Mission at the ESO Headquarters.
This research has made use of the SIMBAD database, operated at CDS, Strasbourg, France. 
\end{acknowledgements}

\bibliographystyle{aa} 
\bibliography{ref}

\clearpage \onecolumn
\longtab{1}{
\begin{longtable}{rllrrrrcccc}
\caption{\label{tab:obs}Program stars. $N$ is the number of spectra entering the mean. (1) \citet{allendep+2004}; (2) \citet{sousa+2008}; (3) \citet{gonzalez+2010}; (4) \citet{bruntt+2010}; (5) \citet{cenarro+2007}; (6) \citet{daSilva+2006}; (7) \citet{sanchezb+2006}; (8) \citet{luck+2007}; (9) \citet{santos+2004}; (10) \cite{mcwilliam1990}; (11) \citet{hekker+2007}; (12) \citet{abia+1988}.}\\
\hline\hline
HD    &  Name      &  Spec.  &  $N$ &  SNR &  $T_\mathrm{eff}$  &  $\log g$  & $[Fe/H]$  & S-index & $\log R'_{HK}$ & Ref.\\
      &            &  Type   &     &   &   [K]  & [cgs] & dex &   & dex  &  \\     
\hline
\endhead
      & Sun       &  G2\,V          & 86  & 850 & 5777 & 4.44 & 0.00 & 0.218 & -4.783 & 1  \\
1461    & \object{HR 72}   &  G0\,V         & 38 & 1500 & 5765 & 4.38 & 0.19 & 0.159 & -5.015 & 2  \\
4915  & \object{HD 4915}   &  G0          & 32   & 1850 & 5638 & 4.52 & -0.21 & 0.209 & -4.782 & 2 \\
7449  & \object{HD 7449}   &  F9.5\,V        & 29  &  1800 & 6024 & 4.51 & -0.11& 0.178 & -4.780 & 2  \\
10700  & \object{\tauCet}    &  G8.5\,V        & 438    & 1275 & 5310 & 4.44 & -0.52 & 0172 & -4.937 & 2  \\
16417  & \object{HR 772}   &  G1\,V          & 22   & 2500 & 5483 & 4.16 & -0.13 & 0.149 & -5.056 & 2   \\
20794 & \object{82 Eri}   &  G8\,V         & 49  & 1000 & 5401 & 4.40 & -0.40 & 0.166 & -4.977 & 2  \\
20807  & \object{HR 1010}    &  G0\,V          & 28 & 1900 & 5866 & 4.52 & -0.23 & 0.166 & -4.879 & 2 \\ 
21019  & \object{HR 1024}   &  G2\,V         & 11  & 2100 & 5468 & 3.93 & -0.45 & 0.150 & -5.066 & 2 \\
21693  & \object{HD 21693}   &  G9\,IV-V        & 21   &  1050 & 5430 & 4.37 & 0.00 &  0.168 & -5.010 & 2 \\
22049  & \object{\epsEri}   &  K2\,V        & 12   &  1000 & 5133 & 4.53 & -0.11 & 0.527 & -4.500 & 2 \\
23249  & \object{\delEri}   &  K1\,III-IV        & 12   &  1000 & 5150 & 3.89 & 0.13 & 0.155 & -5.095 & 2 \\
26967  & \object{\alfHor}   &  K2\,III        & 150   &  300 & 4670 & 2.75 & 0.02 & 0.108 & -5.457 & 2 \\
38858  & \object{HR 2007}   &  G4\,V        & 22   &  1400 & 5733 & 4.51 & -0.22 & 0.173 & -4.925 & 2 \\
38973  & \object{HD 38973}   &  G0\,V        & 15  &  2400 & 6016 & 4.42 & 0.05 & 0.159 & -4.942 & 2 \\
40307  & \object{HD 40307}   &  K2.5\,V        & 15   &  650 & 4977 & 4.47 & -0.31 & 0.198 & -4.996 & 2 \\
50806  & \object{HR 2576}   &  G5\,V        & 14   &  1150 & 5633 & 4.11 & 0.03 & 0.148 & -5.096 & 2 \\
54810  & \object{20 Mon}   &  K0\,III        & 20   &  1200 & 4697 & 2.35 & -0.30 & 0.146 & -5.228 & 2 \\
59468  & \object{HD 59468}   &  G6.5\,V        & 11   &  1350 & 5618 & 4.39 & 0.03 & 0.162 & -5.004 & 2 \\
60532  & \object{HR 2906}   & F6\,IV-V        & 20   &  4000 & 6121 & 3.73 & -0.25 & 0.160 & -4.885 & 3 \\
61421  & \object{Procyon}   & F5\,IV-V        & 70   &  1550 & 6485 & 4.01 & 0.01 & -0.170 & ... & 4 \\
61772  & \object{HR 2959}   & K3\,III        & 20   &  1750 & 3996 & 1.47 & -0.11 & 0.256 & ... & 5 \\
61935 & \object{\alfMon}   & G9\,III        & 20   &  1350 & 4879 & 3.0 & -0.01& 0.134 & -5.266 & 6 \\
65695 & \object{27 Mon}   & K2\,III        & 5   & 725 & 4468 & 2.3 & -0.14 & 0.157 & ... & 6 \\
68978 & \object{HD68978 A}   & G0.5\,V        & 25   & 2200 & 5965 & 4.48 & 0.04 & 0.180 & -4.875 & 2 \\
69830 & \object{HR 3259}   & G8\,V        & 31   & 1050 & 5402 & 4.40 & -0.06 & 0.166 & -5. 008 & 2 \\
72673 & \object{HR 3384}   & G9\,V        & 12   & 1150 & 5243 & 4.46 & -0.41 & 0.180 & -4.930 & 2 \\
73524 & \object{HR 3421}   & G0\,V        & 10   & 1500 & 6017 & 4.43 & 0.16 & 0.154 & -4.982 & 2 \\
73898 & \object{\zetPyx}   & G4\,III        & 40   & 950 & 5030 & 2.80 & -0.43 &0.115 & -5.294 & 7 \\
81169 & \object{\lamPyx}   & G7\,III        & 5   & 1200 & 5119 & 3.13 & -0.07 &0.142 & -5.148 & 8 \\
81797 & \object{\alfHya}   & K2\,II-III        & 5   & 400 & 4186 & 1.7 & 0.00 &0.212 & ... & 6 \\
84117 & \object{HR 3862}   & F8\,V        & 25   & 1750 & 6167 & 4.35 & -0.03 & 0.178 & -4.824 & 9 \\
85444 & \object{39 Hya}   & G6-G8\,III        & 5   & 3000 & 5000 & 2.93 & -0.14 & 0.245 & -4.853 & 10 \\
85512 & \object{HD 85512}   & K6\,V        & 6   & 1450 & 4715 & 4.39 & -0.32 & 0.399 & 4.858 & 2 \\
85859 & \object{HR 3919}   & K2\,III        & 10   & 575 & 4340 & 2.35 & -0.03 & 0.150 & ... & 10 \\
90156 & \object{HD 90156}   & G5\,V        & 4   & 1250 & 5599 & 4.48 & -0.24 & 0.170 & -4.952 & 2 \\
96700 & \object{HR 4328}   & G0\,V        & 8   & 1650 & 5845 & 4.39 & -0.18 & 0.164 & -4.939 & 2 \\
98430 & \object{\delCrt}   & K0\,III        & 15   & 800 & 4580 & 2.35 & -0.43 &0.145 & ... & 11 \\
100407 & \object{\ksiHya}   & G7\,III        & 30   & 1500 & 5045 & 2.76 & 0.21 & 0.212 & -4.947 & 4 \\
102365 & \object{HR 4523}   & G2\,V        & 60   & 1100 & 5629 & 4.44 & -0.29 & 0.187 & -4.792 & 2 \\
102438 & \object{HR 4525}   & G6\,V        & 37   & 1400 & 5560 & 4.41 & -0.29 & 0.170 & -4.951 & 2 \\
102870 & \object{\betVir}   & F8\,V        & 28   & 1850 & 6050 & 4.01 & 0.12 & 0.181 & -4.823 & 4 \\
114613 & \object{HR 4979}   & G3\,V        & 75   & 850 & 5729 & 3.97 & 0.19 & 0.165 & -4.949 & 2 \\
115617 & \object{61 Vir}   & G7\,V        & 36   & 900 & 5558 & 4.36 & -0.02 & 0.167 & -4.984 & 2 \\
115659 & \object{\gamHya}   & G8\,III        & 15   & 950 & 5110 & 2.90 & 0.03 &0.133 & -5.203 & 11 \\
123123 & \object{\piHya}   & K1\,III-IV        & 21   & 725 & 4670 & 2.65 & -0.16 & 0.141 & -5.355 & 11 \\
128620 & \object{\alfCenA}   & G2\,V        & 982   & 875 & 5745 & 4.28 & 0.22 & 0.160 & -4.948 & 4 \\
128621 & \object{\alfCenB}   & K1\,IV        & 6   & 825 & 5145 & 4.45 & 0.30 &0.232 & -4.820 & 4 \\
130952 & \object{11 Lib}   & G9\,III        & 30   & 2000 & 4820 & 2.91 & -0.38 &  0.148 & -5.180 & 10 \\
134060 & \object{HR 5632}   & G0\,V        & 44   & 2050 & 5966 & 4.43 & 0.14 & 0.158 & -4.980 & 2 \\
134606 & \object{HD 134606}   & G6\,IV        & 8   & 1500 & 5633 & 4.38 & 0.27 & 0.150 & -5.070 & 2 \\
134352 & \object{HR 5699}   & G4\,V        & 58   & 2000 & 5664 & 4.39 & -0.34 & 0.167 & -4.968 & 2 \\
136713 & \object{HD 136713}   & K2\,V        & 28   & 750 & 4994 & 4.45 & 0.07 &0.311 & -4.750 & 2 \\
144585 & \object{HR 5996}   & G1.5\,V        & 17   & 1500 & 5914 & 4.35 & 0.33 &0.148 & -5.042 & 2 \\
147675 & \object{\gamAps}   & G8\,III        &  25   & 1800 & 5050 & 3.50 & 0.05 & 0.248 & -4.834 & 12 \\
154962 & \object{HR 6372}   & G6\,IV-V        &  5   & 2300 & 5827 & 4.17 & 0.32 & 0.139 & -5.158 & 2 \\
160691 & \object{\muAra}   & G3\,IV-V        &  279   & 400 & 5780 & 4.27 & 0.30 & 0.149 & -5.101 & 2 \\
162396 & \object{HR 6649}   & F9\,V        &  23   & 2000 & 6090 & 4.27 & -0.35 & 0.152 & -4.992 & 2 \\
171990 & \object{HR 6994}   & F8\,V        & 11   & 3000 & 6045 & 4.14 & 0.06 &0.142 & -5.078 & 2 \\
176986 & \object{HD 176986}   & K2.5\,V        & 28   & 600 & 5018 & 4.45 & 0.00 &0.264 & ... & 2 \\
179949 & \object{HR 7291}   & F8\,V        &  8   & 2400 & 6287 & 4.54 & 0.21 & 0.192 & -4.747 & 2 \\
190248 & \object{\delPav}   & G8\,IV        &  17   & 1300 & 5604 & 4.26 & 0.33 & 0.152 & -5.070 & 2 \\
207129 & \object{HR 8323}   & G2\,V        &  42   & 1050 & 5937 & 4.49 & 0.00 & 0.199 & -4.795 & 2 \\
209100 & \object{\epsInd}   & K5\,V        &  6   & 500 & 4754 & 4.45 & -0.20 & 0.419 & -4.767 & 2 \\
210918 & \object{HR 8477}   & G2\,V        &  12   & 1600 & 5755 & 4.35 & -0.09 & 0.160 & -4.985 & 2 \\
215456 & \object{HR 8658}   & G0.5\,V        &  29   & 2000 & 5789 & 4.10 & -0.09 & 0.146 & -5.068 & 2 \\
221420 & \object{HR 8935}   & G2\,IV-V        &  9   & 2150 & 5847 & 4.03 & 0.33 & 0.135 & -5.188 & 2 \\
\hline
\end{longtable}
}
\end{document}